\documentclass{article}

\usepackage{comment}

\usepackage{spconfa4,amsmath,graphicx}
\usepackage{multirow}
\usepackage{adjustbox}
\usepackage{tabularx,booktabs}
\usepackage{url}
\usepackage{caption}
\usepackage{gensymb}

\newcommand{\C}[1]{}
\usepackage{bm}
\usepackage{bbm}
\usepackage{mathrsfs}
\usepackage{amsmath, amssymb}

\newcommand\blankfootnote[1]{%
  \let\thefootnote\relax\footnotetext{#1}%
  \let\thefootnote\svthefootnote%
}

\DeclareMathOperator{\arctantwo}{arctan2}

\renewcommand*{\vec}[1]{\bm{#1}}

\newcommand*{\fbin}{k} 

\newcommand*{\frm}{l}
\newcommand*{\micnum}{M}
\newcommand*{\ch}{m}

\newcommand*{\stftbin}{\frm, \fbin}

\newcommand*{\srcnum}{J}
\newcommand*{\src}{j}

\newcommand{\tMic}{y}
\newcommand{\fMic}{\MakeUppercase{\tMic}}

\newcommand{\tSrcDry}{s}
\newcommand{\fSrcDry}{\MakeUppercase{\tSrcDry}}

\newcommand{\tNs}{v}
\newcommand{\fNs}{\MakeUppercase{\tNs}}

\newcommand{\doa}{\phi}

\newcommand*{\feat}{\mathcal{F}}
\newcommand*{\Out}{\mathcal{O}}

\newcommand*{\featIn}{\feat_{in}}

\newcommand*{\OutMask}{\Out_{\Mask}}
\newcommand*{\OutRe}{\Out_{\Re}}
\newcommand*{\OutIm}{\Out_{\Im}}

\newcommand*{\Mask}{\mathcal{M}}

\newcommand*{\est}[1]{\widehat{#1}}

\newcommand*{\refr}{\text{ref}} 

\newcommand{\perm}{\pi}

\newcommand{\phase}{\psi}

\title{Location-based Training with Complementary Folded Linear Orderings for Multichannel Speech Separation}
\name{Kaixuan Yang, Stijn Kindt, Nilesh Madhu\thanks{This research was supported by the imec.icon Project UBIWAU, funded by imec and Flanders Innovation \& Entrepreneurship. Microsoft Copilot (corporate license) was used for language editing. The authors are fully responsible for the technical content.}}
\address{\textit{IDLab, Ghent University-imec}, Ghent, Belgium \\
kaixuan.yang@ugent.be, stijn.kindt@ugent.be, nilesh.madhu@ugent.be 
}

\begin{document}
\ninept
\sloppy
\maketitle
\begin{abstract}

Location-based training (LBT) effectively resolves the output permutation problem in multichannel speech separation by imposing deterministic spatial orderings. For planar microphone arrays, LBT typically adopts circular azimuth ordering to cover the full spatial range. However, the resulting cyclic topology introduces a discontinuity at the wrap-around point, increasing learning complexity and limiting the effective use of spatial cues.
This work investigates this limitation by introducing location-based training with folded linear orderings (LBT-FLOs), which collapse circular azimuths into controlled linear orderings. While individual LBT-FLOs exhibit front–back ambiguity, each provides enhanced spatial discriminability over specific azimuth regions. Exploiting their complementarity, we propose an ensemble-style framework that selects among multiple LBT-FLOs using azimuth-guided scoring. Experiments across planar array geometries and reverberant conditions show modest but consistent improvements over circular-ordering LBT, with robustness to azimuth estimation errors.
\end{abstract}
\vspace{-8pt}
\section{Introduction}
\label{sec:intro}
\vspace{-4pt}

In realistic acoustic environments, interfering speech and background noise substantially reduce the target's speech intelligibility and quality. Downstream applications, including hearing-assistive devices and voice-controlled interfaces, need a way to deal with these challenges. Speech separation aims to resolve this by recovering individual sources from such mixtures, thereby supporting both human listening and machine processing.

Deep learning has driven major progress in monaural speech separation, but training remains challenged by the output permutation ambiguity, whereby the correspondence between network outputs and target speakers is undefined. Permutation Invariant Training (PIT) and its variants \cite{PIT,uPIT,CuPIT,FLA_PIT,Prob-PIT} address this issue by dynamically selecting the optimal permutation between estimated outputs and reference sources during training. Single-channel settings separate the speakers by leveraging discriminative acoustic cues such as speaker onsets and vocal characteristics. In contrast, multichannel speech separation systems can additionally exploit spatial cues captured by microphone arrays, leading to more accurate separation performance \cite{CombFeatSS}. Location-based Training (LBT) \cite{LBTconf,LBTjournal} leverages explicit spatial priors to determine the output permutation, outperforming PIT-based approaches in multichannel scenarios.

\begin{figure}[t]
    \centering
    \includegraphics[width=0.9\columnwidth]{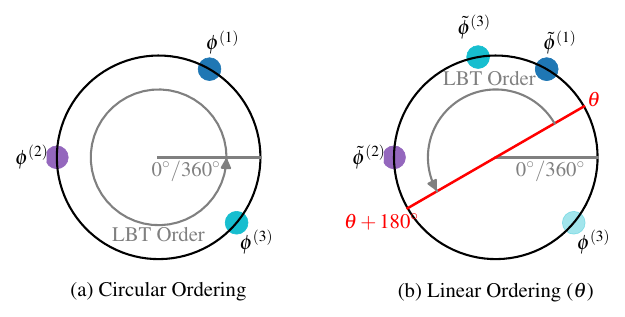}
    \caption{Illustration of circular and linear ordering topologies in LBT for a three‑speaker example. Circular ordering preserves the full azimuthal cycle, whereas linear ordering folds azimuths onto a $180^\circ$ half‑plane along orientation $\theta$ due to front–back ambiguity, causing the source of $\doa^{(3)}$ to be aliased to $\tilde{\doa}^{(3)}$.}
    \label{fig:AzimuthProjection}
\end{figure}

LBT's output permutation is determined by the microphone array geometry. For linear arrays, the well‑known front–back ambiguity causes source directions to collapse onto a $180\degree$ angular range, within which a monotonic ordering of sources can be naturally defined. In contrast, planar arrays inherently support circular ordering that spans the full‑azimuthal range; however, imposing this structure as a learning target results in a discontinuity at the $360\degree/0\degree$ boundary (illustrated in Figure~\ref{fig:AzimuthProjection}), which may impose an additional learning burden, degrading separation performance.

\C{In contrast, planar arrays geometrically support circular ordering to represent sources over the full azimuth range, thereby providing more discriminative spatial guidance for speech separation. Nevertheless, LBT with circular ordering (LBT-CO) enforces a wrap-around at the $360^\circ/0^\circ$ boundary, which introduces an artificial order flip in the learned ordering manifold and, by definition, disrupts spatial consistency.}

\C{With a linear array, the ordering can only be done in a $180\degree$ range due to the well-known front--back ambiguity. This inherently limits the separation performance. Planar arrays support circular ordering (CO), where the full $360\degree$ can be used. However, this results in confusion and discontinuity at the $360\degree/0\degree$ point. This paper hypothesises that this challenge degrades the separation performance.

This paper investigates this effect by forcing a linear LBT ordering onto planar arrays, which we term Folded Linear Ordering (FLO). This obviously has the major drawback of unnecessarily introducing the front--back ambiguity. However, this paper shows that FLO shows better separation in the discriminatory region, certainly around the $360\degree/0\degree$ point. In-depth analysis is performed to highlight this.

To avoid the front--back ambiguity, this paper additionally proposes the use of complementary LBT-FLO networks, where the linear fold is performed along different axes. An azimuth-guide score-based selection criterion picks the network with the optimal axis for separation. Evaluation of LBT-FLO with LBT-CO is performed in reverberant conditions, for two array geometries and several different azimuth combinations.}

We investigate collapsing the circular ordering into controlled linear azimuth axes and implementing LBT with Folded Linear Orderings (LBT-FLOs). Examining a set of LBT-FLOs provides a controlled means to show that the separation network compromises spatial discriminability to accommodate a circular ordering. Based on this, we propose a novel framework that combines complementary LBT-FLOs (CLBT-FLOs) using an azimuth-guided score-based selection policy. This framework is further evaluated against the Circular Ordering baseline (LBT-CO) across two planar array geometries and a wide range of azimuth configurations in a highly reverberant setting.
Collectively, this work offers new insights into ordering-induced trade-offs in LBT and introduces a robust ensemble‑style framework for multichannel speech separation.
\vspace{-0.2cm}

\begin{figure*}[t]
    \centering
    \includegraphics[width=2\columnwidth]{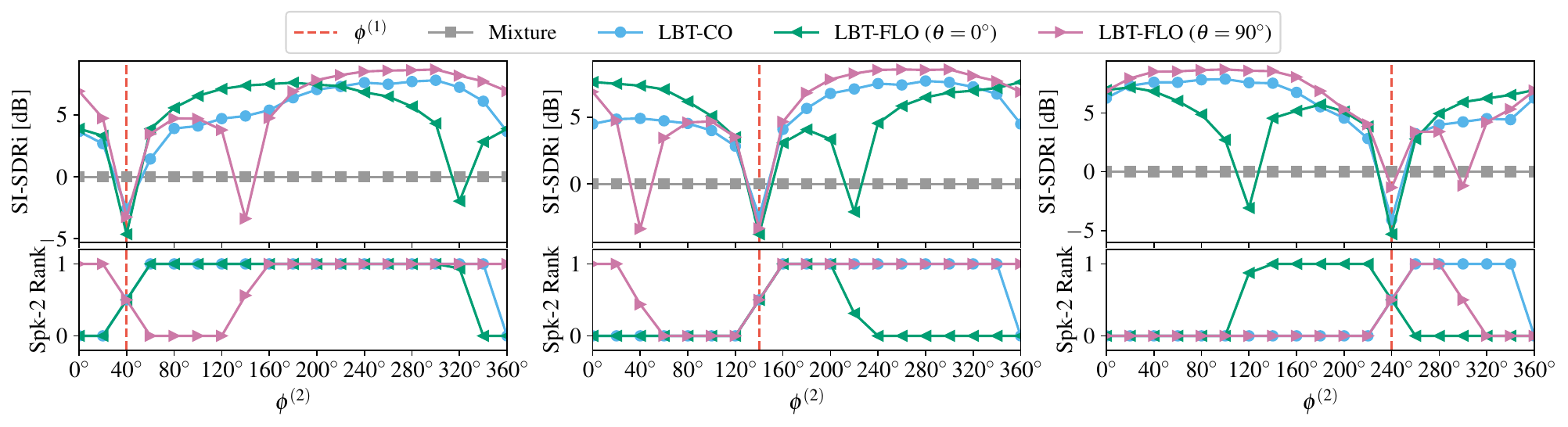}
    \caption{Speech separation performance measured by SI-SDRi and learned output ordering for circular ordering (LBT-CO) and folded linear orderings (LBT-FLOs) with orientations $\theta=0^\circ$ and $\theta=90^\circ$ under anechoic conditions. All LBT models share a common architecture (Section~\ref{sec:exp_settings}).
    Two-speaker mixtures are evaluated by fixing one speaker at $\{40^\circ,\,140^\circ,\,240^\circ\}$ (dotted line) while sweeping the second speaker across all azimuths (x-axis).
    LBT-FLOs show symmetric performance notches around their folding axes due to front--back ambiguity but maintain consistent ordering, whereas LBT-CO exhibits degradations near the $0^\circ/360^\circ$ discontinuity.
    Although no single LBT-FLO is uniformly superior, for every azimuth pair at least one LBT-FLO outperforms LBT-CO.}
    \label{fig:OrderDiscontinuity}
\end{figure*}

\vspace{-2pt}
\section{Signal Model and Location-based Training}\label{sec:signal_LBT}
\vspace{-4pt}
\subsection{Multi‑Channel Signal Model}
\vspace{-4pt}
We assume a compact microphone array with $\micnum$ microphones captures audio signals from $\srcnum$ concurrent static speakers. In the short-time Fourier transform (STFT) domain, the signal at microphone $\ch$ at time frame $\frm$ and frequency bin $\fbin$ is denoted as:
\vspace{-6pt}
\begin{equation}\label{eq:sig}
\fMic_{\ch}(\stftbin) = \sum_{\src=1}^{\srcnum} \fSrcDry_{\ch}^{(\src)}(\stftbin) + \fNs_{\ch}(\stftbin) \, ,
\end{equation}
where $\fSrcDry_{\ch}^{(\src)}(\stftbin)$ represents the direct-path speech signal from speaker $\src$ to microphone $\ch$, and $\fNs_{\ch}(\stftbin)$ accounts for the combined effects of reverberation from all $\srcnum$ sources and background noise.
The objective of speech separation is to estimate the individual dry speech signals of each speaker at a designated reference microphone, denoted by $\hat{\fSrcDry}_{\refr}^{(\src)}$. 
For simplicity, the reference microphone index~$\refr$ is omitted when referring to $\fSrcDry$ and $\hat{\fSrcDry}$ henceforth. Boldface symbols denote vectors, defined either along the microphone index or the source index, depending on the context. For example, $\Vec{\fMic} = [\fMic_{1}, \fMic_{2}, \cdots, \fMic_{\micnum}]^T$, $\Vec{\hat{\fSrcDry}} = [\hat{\fSrcDry}^{(1)}, \hat{\fSrcDry}^{(2)}, \cdots, \hat{\fSrcDry}^{(\srcnum)}]^T$.

\vspace{-4pt}
\subsection{Location-based Training}
\vspace{-2pt}

In multi-speaker speech separation, a primary challenge is resolving the output permutation ambiguity among speakers to ensure a consistent and stable training process. 
Location‑based training (LBT) \cite{LBTconf, LBTjournal} addresses this challenge by defining a two‑fold learning target: estimating the separated speech signals while simultaneously enforcing a deterministic output permutation $\pi$, determined by speaker azimuths or distances relative to the microphone array. Prior work has shown that azimuth-based ordering outperforms both distance-based ordering and permutation-invariant training (PIT) \cite{LBTjournal}. 

Effectively, azimuth-based LBT defines speaker order with reference to an {\bfseries orientation azimuth} $\theta$. 
For a planar array,  $\theta = 0^\circ$ is conventionally used as the orientation azimuth, corresponding to a reference direction in the source-array coordinate system.
Let the azimuth angles of the $\srcnum$ speakers be denoted by $\doa^{(1)}, \doa^{(2)}, \ldots, \doa^{(\srcnum)} \in [0^\circ, 360^\circ)$.
In this case, the {\bfseries ordering coordinate} directly corresponds to the speaker azimuth itself, and a {\bfseries Circular Ordering} $\perm_{C}$ over the full $360^\circ$ azimuth range is defined such that \( \doa^{\perm_{C}(1)} < \doa^{\perm_{C}(2)} < \cdots < \doa^{\perm_{C}(\srcnum)} \).
For a linear array, the inherent front–back ambiguity renders sources at symmetric azimuths across the array axis indistinguishable, collapsing the full azimuth range onto a $ 180^\circ $ half‑plane. As a result, the array axis defines an orientation azimuth $\theta \in [0^\circ, 180^\circ)$ and an ordering coordinate system which spans a reduced azimuth domain $[\theta,\, \theta + 180^\circ]$. The ordering coordinate for each speaker $\src$ is obtained by projecting its azimuth as follows:
\begin{equation}\label{eq:azi_proj}
\tilde{\doa}^{(\src)}
= \theta
+ \left|
\operatorname{mod}\!\left(\doa^{(\src)} - \theta + 180^\circ,\; 360^\circ\right)
- 180^\circ
\right|.
\end{equation}
Consequently, \textbf{Linear Ordering} along orientation~$\theta$ is specified by a permutation $\perm_{L_\theta}$ such that
\( \tilde{\doa}^{\perm_{L_\theta}(1)} < \tilde{\doa}^{\perm_{L_\theta}(2)} < \cdots < \tilde{\doa}^{\perm_{L_\theta}(\srcnum)} .\) 
Figure~\ref{fig:AzimuthProjection} illustrates ordering in circular and linear topologies for an example 3-speaker scenario.
\C{Figure~\ref{fig:AzimuthProjection} visualises the circular and linear ordering topologies in LBT for a three‑speaker example. It also shows that linear ordering can be enforced on planar arrays. However, from the perspective of front--back ambiguity, circular ordering might better match the planar array geometries. Nevertheless, there are also some limitations to circular ordering.}
\vspace{-0.2cm}

\subsection{Limitations of Conventional LBT}
\vspace{-4pt}

Conventionally, planar-array LBT adopts circular ordering, as it provides a rich ordering coordinate system spanning the full azimuth range. However, enforcing a wrap‑around at the orientation azimuth introduces a discontinuity in the ordering manifold: directions just below $360^\circ$ and just above $0^\circ$ are mapped to opposite ends of the ordering, so small angular changes near this boundary can cause abrupt order flips. This behaviour increases learning complexity and can divert model capacity away from accurate signal reconstruction.

Inspired by linear‑array LBT, folding ordering coordinates across a controlled orientation azimuth $\theta$  yields a continuous and monotonic linear ordering. In particular, when source azimuths cross the orientation axis, their ordering coordinates are folded back into a bounded range $[\theta,\theta + 180^\circ]$ rather than wrapping around. Motivated by this observation, we propose {\bfseries Folded Linear Ordering} (LBT‑FLO) to study the impact of ordering topology on separation training, and compare it against LBT with circular ordering (LBT‑CO).

Figure~\ref{fig:OrderDiscontinuity} compares LBT-FLO along two different folding axes with LBT-CO under anechoic conditions. The dotted vertical line indicates the position of the first speaker, while the second speaker's location is indicated on the x-axis. Separation performance is measured using Scale-invariant Signal-to-Distortion Ratio improvement (SI-SDRi) \cite{SI-SDR}. Additionally, the output channel of the second speaker is plotted (Spk-2 Rank), indicating the learned permutation. Since the ordering is neural architecture independent, all methods adopted a common architecture, described in detail in Section~\ref{sec:exp_settings}.

LBT‑FLOs exhibit symmetric performance notches around the array orientation as a result of front–back ambiguity. However, when considered collectively, there always exists a best‑performing LBT‑FLO that outperforms LBT‑CO across all azimuth configurations. This observation provides empirical support for our hypothesis that circular ordering diverts model capacity toward enforcing an artificial topology, thereby limiting its effectiveness for speech separation.

\begin{figure}[ht]
    \centering
    \includegraphics[width=1.0\columnwidth]{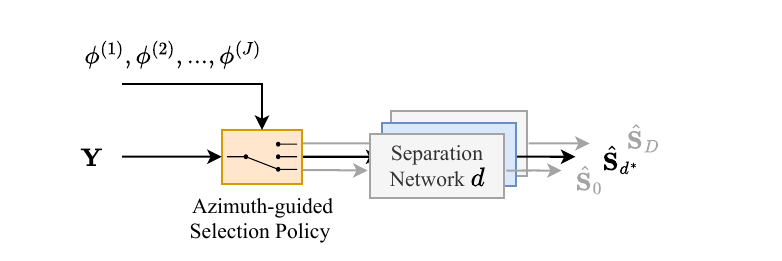}
    \caption{Diagram of the proposed ensemble‑style framework illustrating selection among complementary LBT‑FLO separation networks.}
    
    \label{fig:Diagram}
\end{figure}

\vspace{-12pt}
\section{Proposed Method} \label{sec:proposed_method}
\vspace{-4pt}

Inspired by the observations above, we propose a novel ensemble‑style framework that combines complementary LBT‑FLOs (CLBT-FLOs) through a selection policy to achieve robust speech separation. Specifically, for each mixture, a single LBT‑FLO is selected and applied based on the estimated spatial configuration, rather than evaluating multiple networks in parallel. An overview of the proposed approach is illustrated in Figure~\ref{fig:Diagram}.

The network selection policy (indexed by $d$) is designed based on the observation that the separation performance strongly depends on how close the ordering coordinates are. This effect is quantified through a local‑crowding score defined as follows:
\vspace{-4pt}
\begin{equation}\label{eq:dist_speakers}
C_d =
\Bigg(
    \sum_{\src=1}^{\srcnum} \sum_{\src'=1}^{\src-1} \max\!\Big(| \tilde{\doa}^{(\src')}_d - \tilde{\doa}^{(\src)}_d |,\; \varepsilon \Big)^{-p}
\Bigg)^{-1/p}\,,
\end{equation}
where $\varepsilon=1$ prevents numerical instability, $p$ is set to 4.

In addition, as separation gains tend to diminish near the boundaries of linear orderings, source pairs are more effectively separated when positioned farther away from these boundaries. 
To capture this effect, we define a boundary-distance score as follows:
\vspace{-4pt}
\begin{equation}\label{eq:dist_boundary}
B_d = 
\left(
    \sum_{\src=1}^{\srcnum} \max\!\big(\min( \tilde{\doa}^{(\src)}_d - \theta_d,\; \theta_d + 180^\circ -\tilde{\doa}^{(\src)}_d ),\; \varepsilon \big)^{-p}
\right)^{-1/p}
\end{equation}
The final score is computed as: $\mathrm{Score}_d = w_C\, C_d + w_B\, B_d$,
where the weights are empirically set to $w_C = 1.0$ and $w_B = 0.2$. The selected network is then determined by $ d^\star = \arg\max_d \mathrm{Score}_d $.

\section{Experimental Settings} \label{sec:exp_settings}
\vspace{-4pt}
\subsection{Training Data and Settings}
\vspace{-4pt}

\begin{figure}[h!]
    \centering
    \includegraphics[width=0.35\columnwidth]{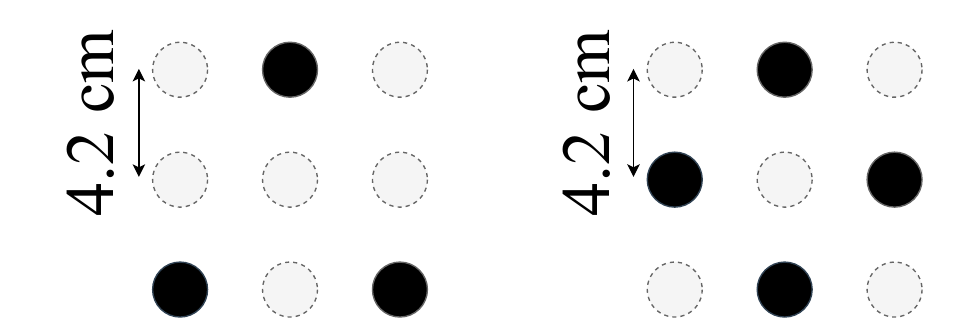} 
\caption{Microphone array configurations positioned on a $3{\times}3$ grid. 
Left: Pseudo–Equilateral Triangle array (PET‑3). 
Right: Uniform, Cross-shaped Rectangular Array (URA‑4+). }
    \label{fig_array_geometries}
\end{figure}

Array signals are generated on a $3{\times}3$ grid as depicted in Figure~\ref{fig_array_geometries}. 
To facilitate robust training, we employ the diverse and well-controlled simulation setup in~\cite{Alexander_LDE}.
Room impulse responses (RIRs) are simulated across 10 distinct room configurations, with reverberation times ($T_{60}$) ranging from 0.20\,s to 0.80\,s. For each room, 7 array center positions and 4 source-array distances are considered, with azimuth angles sampled at a resolution of $5^\circ$. Anechoic RIRs are generated in parallel to provide dry reference signals. 
Clean speech signals are drawn from the TIMIT \cite{TIMIT} and PTDB‑TUG corpora \cite{PTDB-TUG} and segmented into 2‑second utterances, with background noises generated as spatially diffuse and spectrally white.
A total of 40,320 unique acoustic scenarios are generated per epoch, with mixtures of one or two concurrent speakers (\( \srcnum=\{1,2\} \)) positioned at different azimuth angles, signal-to-noise ratios (SNRs) are uniformly sampled in range [0, 30]\,dB. 

\vspace{-4pt}

All audio signals are resampled to a sampling rate of 16\,kHz. 
The Short-Time Fourier Transform (STFT) employs a 512‑sample (32~ms) analysis window, a hop size of 160 samples (10~ms), and square-root Hann windows for both analysis and synthesis to ensure perfect overlap–add reconstruction.

We consider two perpendicular ordering orientations, $\theta = 0^\circ$ and $\theta = 90^\circ$, to provide complementary spatial coverage for the two-speaker $\srcnum = 2$ case. Note: we focus on $\srcnum = 2$ for the proof of concept. When $\srcnum > 2$, LBT-FLOs may collapse sources with widely separated azimuths; nevertheless, they can still function as complementary alternatives when combined with a LBT-CO -- this generalisation is out of current scope. 

\vspace{-4pt}
\subsection{Network Model}
\vspace{-4pt}
The study presented here is not tied to a specific neural architecture. We use the Convolutional Recurrent U-Net for Speech Enhancement (CRUSE) \cite{Eff_CRUSE} as the neural backbone, which comprises a convolutional encoder, a GRU-based bottleneck, and a transposed-convolution decoder with skip connections. Batch normalization and Leaky ReLU activations are applied after each convolutional layer.
The CRUSE network has been easily adapted for multichannel applications by including spatial information into the input features $\featIn(\stftbin) \in \mathbb{R}^{2\micnum+1}$: 
\vspace{-4pt}
\begin{equation}\label{eq:feature}
\featIn(\stftbin) =
\begin{bmatrix}
\cos\{\angle\vec{\fMic}(\stftbin)\}, \sin\{\angle\vec{\fMic}(\stftbin)\}  \\
\log_{10} |\fMic_{\refr}(\stftbin)| - \mathbb{E}_{\stftbin}\left\{ \log_{10} |\fMic_{\refr}(\stftbin)| \right\}
\end{bmatrix},
\end{equation}
where the trigonometric features encode the spatial pattern, the detrended log-magnitude of the reference channel encodes the spectral pattern. The scalar expectation $\mathbb{E}_{t,f}\{\cdot\}$ is implemented as a causal running mean over a 300~ms context. All convolutional kernels are 64-dimensional to capture multi-channel information \cite{Alexander_LDE}.


To estimate the magnitude and phase of separated sources, a hybrid representation $[\vec{\OutMask}(\stftbin), \vec{\OutRe}(\stftbin), \vec{\OutIm}(\stftbin)]^T \in \mathbb{R}^{3\srcnum} $ is adopted for separated sources \cite{LDE_Insights}. The enhancement mask is calculated as: $\vec{\Mask}(\stftbin) = 10^{\vec{\OutMask}(\stftbin)}$ and enhanced phase is obtained from $ \est{\vec{\phase}}(\stftbin) = \arctantwo \left( \vec{\OutRe}(\stftbin), \vec{\OutIm}(\stftbin) \right) $,
where $\vec{\Out}_{\{\Mask,\Re, \Im\}}(\stftbin)\in \mathbb{R}^{\srcnum} $. Finally, separated sources can be obtained by:
\vspace{-4pt}
\begin{equation}\label{eq:est}
\hat{\vec{\fSrcDry}}(\stftbin) = \vec{\Mask}(\stftbin) |\fMic_{\refr}(\stftbin)| e^{\jmath\est{\vec{\phase}}(\stftbin)} \,.
\end{equation}
With location-based label assignment $\perm$, we use a power-compressed complex error loss \cite{Eff_CRUSE}, which can be formulated as:
\vspace{-4pt}
\begin{equation}
\begin{aligned}
\mathcal{E}(\hat{\Vec{\fSrcDry}}, \perm(\Vec{\fSrcDry}) )
&= \frac{1}{ || \lvert \Vec{\fSrcDry} \rvert^{\alpha} ||_2^2 }
\Big[
  \lambda \, ||  \lvert \mathcal{P}\{\hat{\Vec{\fSrcDry}}\} \rvert^{\alpha} - \lvert \perm(\Vec{\fSrcDry}) \rvert^{\alpha} ||_2^2
  \\
&
  + (1 - \lambda) \, 
      || \lvert \mathcal{P}\{\hat{\Vec{\fSrcDry}}\} \rvert^{\alpha} e^{j\angle \mathcal{P}\{\hat{\Vec{\fSrcDry}}\} }
      - \lvert \perm(\Vec{\fSrcDry}) \rvert^{\alpha} e^{j\angle \perm(\Vec{\fSrcDry})} ||_2^2
\Big],
\end{aligned}
\end{equation}
where 
$\mathcal{P}\{\cdot\} = \text{STFT}\{\text{iSTFT}\{\cdot\}\} $ is a backward-forward transform to enhance phase consistency,
with \( \alpha=0.3 \) controlling the power of magnitude compression, \( \lambda = 0.5 \) balancing the contribution of magnitude and phase mapping.
Note $d$ is omitted here for simplicity.

The networks are trained using the AdamW optimiser \cite{AdamW} with a fixed learning rate of \( 8 \times 10^{-5} \) for up to 200 epochs, and the best-performing checkpoints are selected based on validation performance. Model complexities are shown in Table~\ref{tab:complexity_comparison}.


\begin{table}[t]
\centering
\caption{Module‑wise complexity for LBT‑CO and CLBT‑FLOs.}
\label{tab:complexity_comparison}
\begin{adjustbox}{width=0.90\columnwidth}
\begin{tabular}{lcc}
\toprule
\textbf{Module} &
\textbf{Parameters (M)} &
\textbf{MACs (M) per frame} \\
\midrule
Encoder (PET-3 / URA-4+) & 0.08 & 3.01 / 3.10 \\
GRU-based Bottleneck    & 1.39 & 1.34        \\
Skip Connections        & $5\times10^{-4}$ & 0.03 \\
Decoder                 & 0.08 & 2.96        \\
\midrule
Total for LBT‑CO      & 1.55 & 7.34 / 7.44 \\
Total for CLBT‑FLOs (proposed)   & 1.55$\times D$ & 7.34 / 7.44 \\
\bottomrule
\end{tabular}
\end{adjustbox}
\end{table}

\vspace{-2pt}
\subsection{Evaluation Settings}
\vspace{-4pt}
A realistic meeting room with dimensions $7.50 \times 5.00 \times 2.65~\text{m}^3$ ($T_{60}=0.65~s$) is simulated for evaluation. The microphone array is placed at the room center, with a source-array distance of $1~\mathrm{m}$. Speech samples from the TSP dataset \cite{Kabal2002TSP} are concatenated into long utterances ($>8$s) for both male and female speakers. 
Each method is evaluated using $16$ long-speech speaker pairs equally covering all gender combinations. 
For each speaker pair, 6 distinct speaker azimuths are randomly sampled without replacement from  $\{0^\circ,\, 20^\circ,\, \ldots,\, 340^\circ\}$ to form independent reference configurations, while the second speaker varies exhaustively across the same set.

\begin{figure}[h]
    \centering
    \includegraphics[width=\columnwidth]{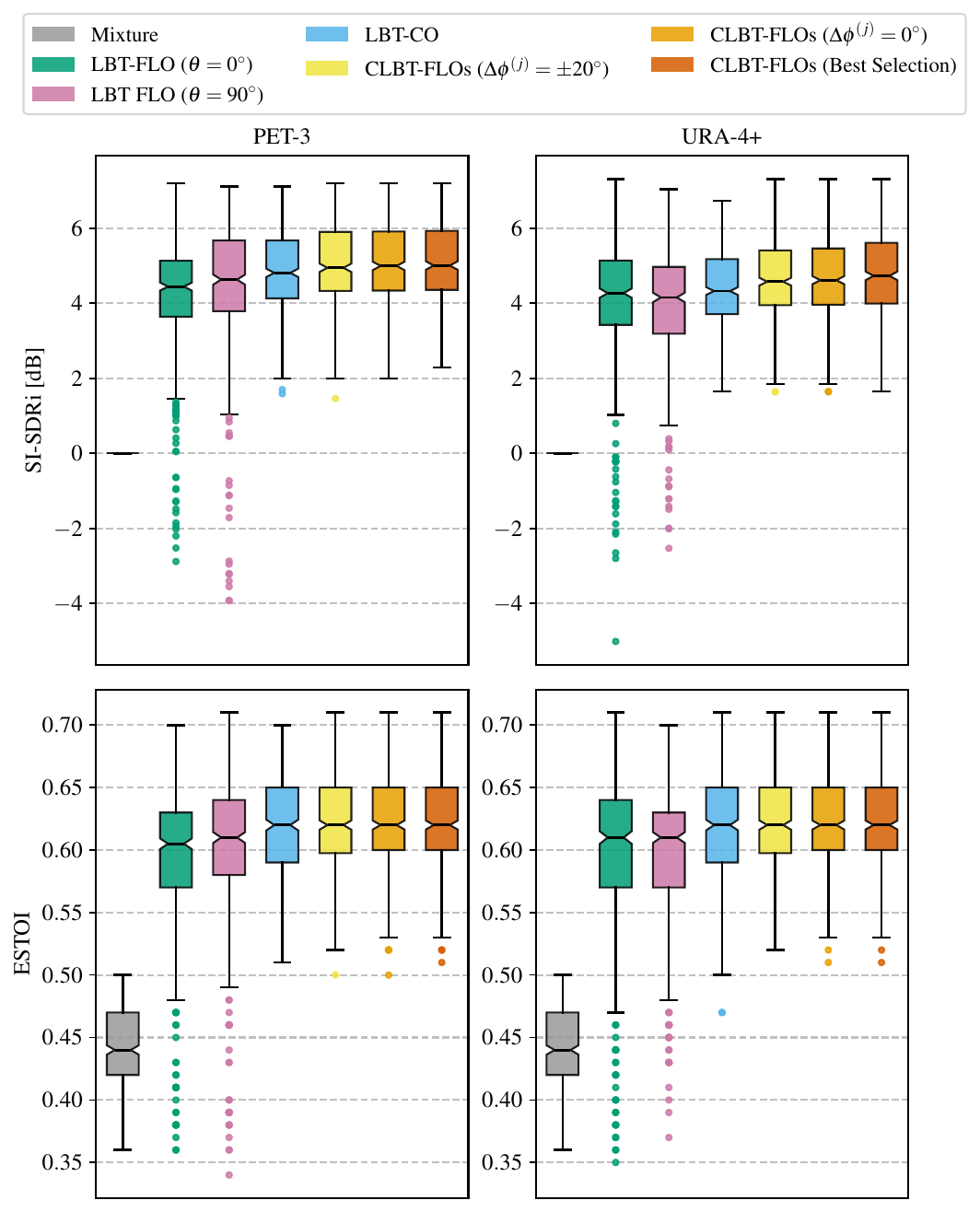}
    \caption{Comparison of speech separation performance across array geometries using SI-SDR and ESTOI. Configurations in which speakers share the same azimuth are excluded from this plot.
    Results are shown for the LBT with circular ordering (LBT-CO), individual folded linear orderings (LBT-FLOs), and the proposed CLBT-FLO framework under oracle ($\Delta\phi^{(j)} = 0^\circ$) and perturbed ($\Delta\phi^{(j)} = \pm20^\circ$) azimuth guidance conditions, as well as the highest‑scoring selection (Best Selection).}
    \label{fig:BoxPlot}
\end{figure}

\vspace{-14pt}
\section{Experimental Results} \label{sec:exp_results}
\vspace{-4pt}
We evaluate speech separation performance using SI-SDR and extended Short-Time Objective Intelligibility (ESTOI) \cite{estoi}, as shown in Figure~\ref{fig:BoxPlot}. 
As a consolidated baseline, LBT‑CO consistently achieves substantial improvements over the input mixture across all conditions, confirming the effectiveness of LBT in resolving speaker permutation ambiguity for planar microphone arrays. In comparison, individual LBT‑FLOs exhibit slightly degraded average performance and long‑tailed failure cases. 
Nevertheless, the proposed selection strategy enables higher spatial discriminability by selecting an appropriate linear ordering for each scenario. As a result, the CLBT-FLOs guided by oracle azimuth ($\Delta\phi^{(j)}=0^\circ$) consistently improves performance relative to individual LBT‑FLOs and LBT-CO. 
Although the average improvements over LBT-CO are modest, they are consistently observed across the evaluation set. This finding is supported by paired Wilcoxon signed-rank tests, yielding effect sizes $r>0.36$, which indicate moderate-to-large effects.

\begin{figure}[ht]
    \centering
    \includegraphics[width=0.90\columnwidth]{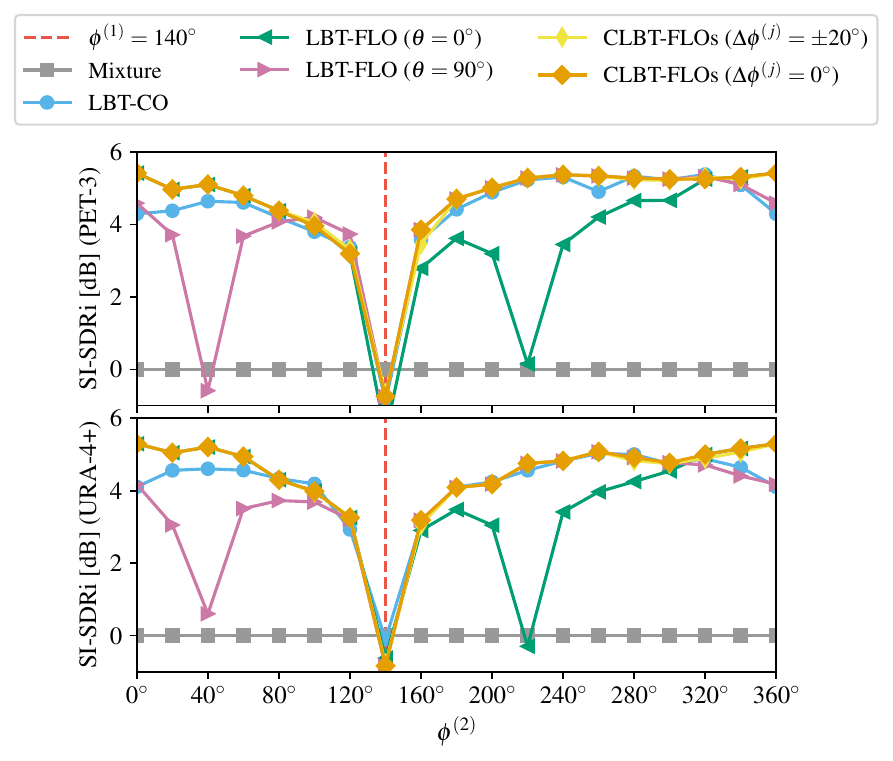} 
    \caption{Proposed selection policy over azimuth pairs for URA‑4+ and PET‑3 arrays, averaged over 16 speaker pairs, demonstrating preference for advantageous LBT‑FLOs.}
    \label{fig:ModelSelection}
\end{figure}

To further assess the robustness of these improvements, we consider two reference conditions. An oracle upper bound is obtained by selecting the highest‑scoring network for each test sample, while a lower‑bound setting is constructed by introducing random azimuth estimation errors of $\pm20^\circ$ per speaker. Notably, the proposed method closely approaches the oracle performance and remains comparable under the lower‑bound condition, demonstrating strong robustness to azimuth estimation errors encountered in realistic acoustic environments. Besides, the PET‑3 array appears to outperform the URA‑4+ array. This behaviour may be related to differences in effective aperture and array geometry; however, a detailed analysis is beyond the scope of this study.
Figure~\ref{fig:ModelSelection} visualizes the selection policy over azimuth pairs, demonstrating consistent preference for advantageous LBT‑FLOs. 
We observe similar trends when the selection policy is evaluated using ESTOI, which are illustrated on the accompanying demo page. To complement the quantitative results, readers are invited to listen to representative speech‑separation audio samples at \url{https://aspire.ugent.be/demos/IWAENC2026KY/}.

\vspace{-20pt}
\section{Conclusion} \label{sec:conclusion}
\vspace{-4pt}
\C{This work identified a key limitation of location‑based training with circular ordering (LBT-CO) for planar microphone arrays, namely the wrap‑around discontinuity that degrades spatial consistency and separation performance. }

We examine a fundamental limitation of location‑based training with circular ordering (LBT‑CO) for planar microphone arrays, where enforcing a cyclic topology hinders the effective use of spatial cues for speech separation. To address this, we propose location‑based training with folded linear orderings (LBT‑FLOs), which enhance spatial discriminability via azimuth folding, and introduce a complementary ensemble selection framework (CLBT‑FLOs). Experiments across array geometries and realistic conditions show modest but consistent gains over circular ordering, with strong robustness to azimuth estimation errors. Future work will extend the framework to multi‑speaker scenarios and parameter‑efficient implementations.

\bibliographystyle{IEEEbib}
\bibliography{references}

\end{document}